\documentclass[10pt, oneside]{article}   	% use "amsart" instead of "article" for AMSLaTeX format
\usepackage{geometry}                		% See geometry.pdf to learn the layout options. There are lots.
\pdfoutput=1
\usepackage{graphicx}				% Use pdf, png, jpg, or eps§ with pdflatex; use eps in DVI mode

\newcommand{\eqb}{\begin{equation}}			% bgn equation
\newcommand{\eqe}{\end{equation}}			% end equation
\newcommand{\fb}[2]{\frac{#1}{#2}} 			% fraction built up
\renewcommand{\d}{\mathrm{d}} 				% tot deriv/inf interval
\newcommand{\fonttherm}{\mathnormal}		% thermal variables font
\newcommand{\xT}{\fonttherm{T}}				% temperature
\newcommand{\xTL}{\fonttherm{T_{L}}}		% liquid temperature
\newcommand{\xTV}{\fonttherm{T_{V}}}		% vapor temperature
\newcommand{\xTC}{\fonttherm{T_{c}}}		% critical temperature
\newcommand{\xlam}{\fonttherm{\lambda}}		% latent heat of vap
\newcommand{\xrhoV}{\fonttherm{\rho_{V}}}	% density of vapor phase
\newcommand{\xsig}{\fonttherm{\sigma}}		% surface tension
\newcommand{\xRc}{\fonttherm{R_{c}}}		% critical radius
\newcommand{\xEc}{\fonttherm{E_{c}}}		% critical energy
\newcommand{\xSH}{\fonttherm{SH}}					% SH
\title{Effective exploitation of a geyser bubble-chamber equipment as a background-free fast neutron detector}

\begin{document}
\maketitle

{\centering

R.~Bertoni$\mathrm{^1}$,
G.~Bruno$\mathrm{^{2,6}}$,
N.~Burgio$\mathrm{^{3,4,5}}$,
M.~Corcione$\mathrm{^{4,5,b}}$,
L.~Cretara$\mathrm{^{4}}$,
M.~Frullini$\mathrm{^{4}}$,\\
W.~Fulgione$\mathrm{^{6,7,a}}$, 
G.~Palmerini$\mathrm{^8}$,
A.~Quintino$\mathrm{^{4,5}}$,
N.~Redaelli$\mathrm{^{1}}$,
A.~Santagata$\mathrm{^{3,4,5}}$,\\
D.~Sorrenti$\mathrm{^{1,9}}$,
L.~Zanotti$\mathrm{^{1,10}}$.\\
}
\vspace{1cm}

{\small

1) INFN, Sezione di Milano Bicocca, P.za della Scienza 3, I-20126 Milano, Italy

2) New York University Abu Dhabi, United Arab Emirates

3) ENEA, Centro Ricerche Casaccia, (Roma), Italy

4) DIAEE, Sapienza Universit\`a di Roma, Via Eudossiana 18, I-00184 Roma, Italy

5) INFN, Sezione di Roma La Sapienza, P.le Aldo Moro 2, I-00185 Roma, Italy

6) INFN, LNGS, Via G. Acitelli 22, I-67100 Assergi (L'Aquila), Italy

7) INAF, Osservatorio Astrofisico di Torino, I-10025 Pino Torinese (Torino), Italy

8) SIA, Sapienza Universit\`a di Roma, Via Salaria 851, I-00138 Roma, Italy 

9) Dip.Inform., Sist. e Comunic., Universit\`a di Milano Bicocca, Italy

10) DISAT, Universit\`a di Milano Bicocca, P.za della Scienza 3, I-20126 Milano, Italy\\

$\mathrm{a)}$  walter.fulgione@lngs.infn.it

$\mathrm{b)}$  massimo.corcione@uniroma1.it
}


%\maketitle

\begin{abstract}

MOSCAB equipment, a geyser-concept bubble-chamber originally thought for the search of dark matter in the form of WIMPs, is employed for the detection of fast neutrons. 
Once the background-free operating conditions are determined such that the detector is sensitive only to neutrons, which occurs when the neutron energy threshold required for nucleation is higher than approximately 2.5 MeV, the detector response to fast neutrons is investigated using a $\mathrm{^{241}}$AmBe neutron source. 

Sets of detection efficiency functions are then produced via Monte Carlo simulations and post-processing, their validation being performed experimentally and discussed. 
Finally, the use of the detector to measure the fast neutron activity of very weak n-sources in clean environments, as well as to monitor the cosmic ray variations through the neutron component of the Extensive Air Showers, is considered.
\end{abstract}

\section{Introduction}
\label{intro}

The MOSCAB bubble-chamber detector relies on the geyser technique, a variant of the superheated liquid technique of extreme simplicity originally introduced by Hahn and Reist to detect fissions fragments \cite{000}.  
It was thought, designed, built and developed by Antonino Pullia and his team, to search for dark matter in the form of WIMPs through a possible spin-dependent interaction occurring with fluorine nuclei, as illustrated and discussed by Bertoni et al. \cite{0000} and Antonicci et al. \cite{001}. 
On the other hand, the same equipment, or a simplified version of the same device, can be effectively operated to detect fast neutrons, which is the topic of the present paper.

MOSCAB detector has been firstly tested above ground at Milano-Bicocca University using, as a target liquid, 0.7L  of octafluoropropane ($\mathrm{C_3F_8}$) kept at different metastability conditions. Later, the apparatus was moved underground to the INFN Gran Sasso National Laboratories (LNGS), repeating both background and neutron sensitivity measurements, which were performed for two different configurations of the detector, equipped with either a 2L vessel filled with 1.2L of $\mathrm{C_3F_8}$ or a 18L vessel filled with 13L of $\mathrm{C_3F_8}$. 

The operation principles of MOSCAB detector, exactly as for any traditional bubble chamber, are based on the theoretical model originally proposed by Seitz \cite{0001}, according to which pressure and temperature of the metastable liquid determine the minimum recoil energy required for bubble nucleation, also called the critical energy. 
On the other hand, since the critical energy must be released within a sufficiently small volume of the sensitive liquid, bubble nucleation depends also on the stopping power of the ionizing particle, which makes the detector entirely unaffected by the backgrounds due to recoiling electrons and minimum ionizing radiation, provided that the critical energy is not too low. 
This means that, besides neutrons, the only residual particle-induced background is represented by $\alpha$-decay events, yet, as it will be shown, such a source of background can be completely muted by operating the detector above specific recoil energy thresholds.

In this general framework, thanks to the extremely low neutron flux typical of the environmental background of LNGS underground laboratories, as well described by Wulandari et al.\cite{002}, we could easily conduct measurements of the residual internal background of the detector.
% \textcolor{red}{along with the evidence that the detector maintains its sensitivity to neutron recoils}. 
Subsequently, the bubble nucleation model describing the behaviour of the detector operated in different configurations and exposed to a neutron source is presented, and the related detection efficiency functions
%response functions \textcolor{red}{of the detector} 
for fast neutrons with energies up to 20 MeV are employed to compare the experimental data with the numerical results of Monte Carlo simulations and post-processing.
Finally, the detector sensitivity and its possible applications in measuring and monitoring low fluxes of fast neutrons in the absence of background are also discussed.

\section{Basic outlines of the detector }
\label{sec:2}

The MOSCAB detector basically consists of a closed quartz vessel filled with a target liquid, i.e., $\mathrm{C_3F_8}$, and its saturated vapour. 
The bottom of the vessel contains the sensitive liquid kept at a constant temperature $\xTL$ by an external thermal bath, whereas the saturated vapour located on top of it is maintained at a lower constant temperature $\xTV$. 
Considering that the pressure exerted inside the vessel is the saturation pressure ${\fonttherm{p_s(T_{V})}}$ imposed by the vapour, which is lower than the equilibrium saturation pressure $\fonttherm{p_s(T_{L})}$, the target liquid is in underpressure, and then in a superheated state, whose degree of metastability can be identified using the reduced superheat parameter $\xSH$ introduced by d'Errico \cite{004}
\eqb
\xSH=\fb{\xTL-\xTV}{\xTC-\xTV}  \label{eq:1}
\eqe
where $\xTC$ is the critical temperature of the target liquid, i.e.,  $\xTC$ $\fonttherm{\mathrm{= 71.87~^{o}C}}$. 

The degree of metastability is continuously monitored by measuring both the liquid temperature $\xTL$ and the saturated vapour pressure $\fonttherm{p_s(T_{V})}$ with a sampling time interval of 6 seconds, then calculating the vapour temperature $\xTV$ using the fluid properties at saturation extracted from the NIST Chemistry Web Book \cite{005}.

\begin{figure} % figuur 1
\begin{minipage}{\columnwidth}
\centering
\resizebox{0.8\textwidth}{!}{%
  \includegraphics{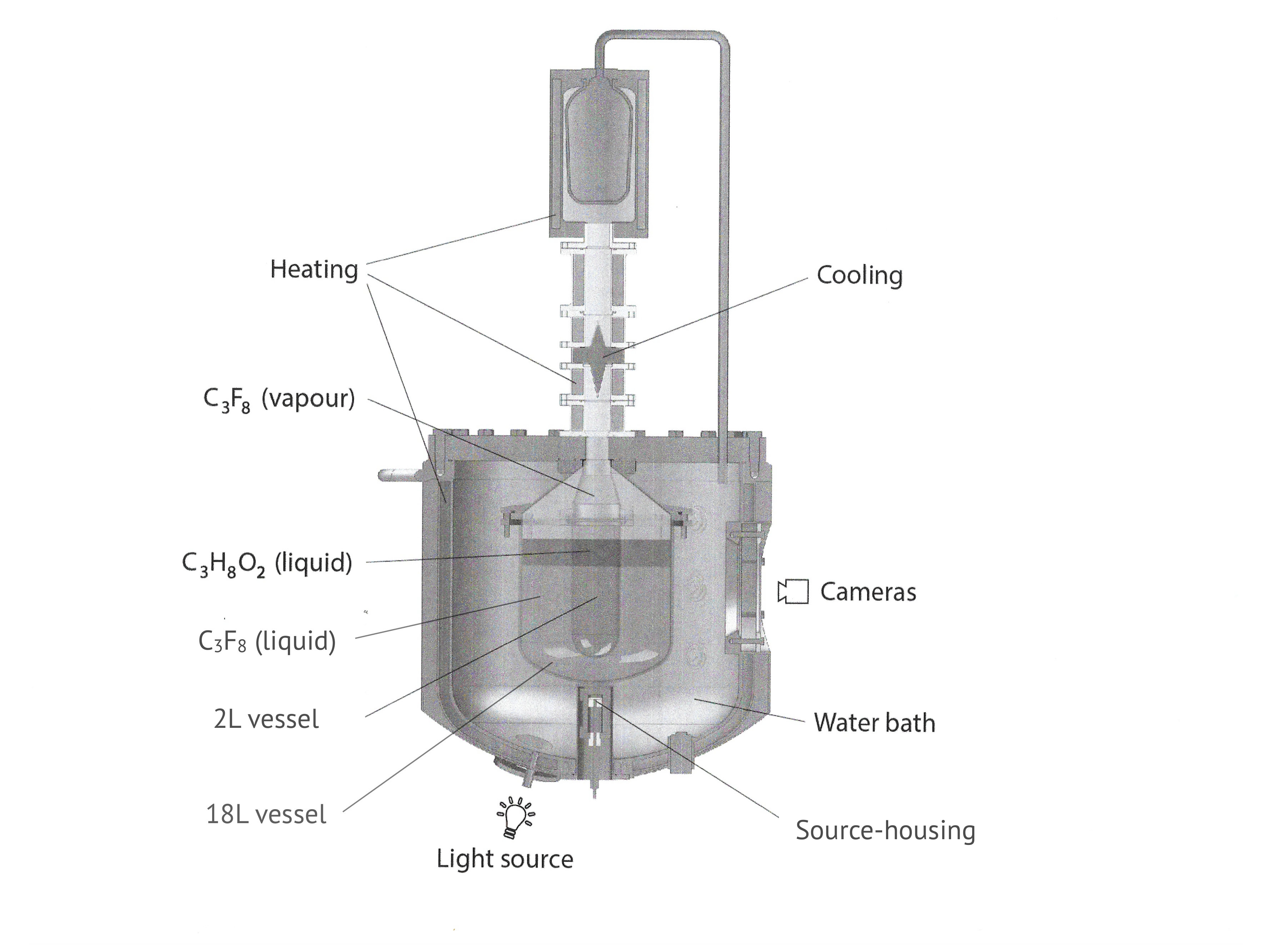}}
%\framebox[\columnwidth][c]{\raisebox{0pt}[20mm][20mm]\includegraphics{Fig_1bw.pdf}}
\end{minipage}
\caption{The MOSCAB bubble chamber. Both the 2L and 18L vessels are reported.}
\label{fig.1}
\end{figure}

A  schematic view of the MOSCAB bubble chamber is displayed in Figure  \ref{fig.1}, in which the two different configurations of the detector discussed in the present work, one equipped with a 2L vessel and the other with a 18L vessel, are shown superimposed. 
A detailed description of the apparatus can be found in the study performed by Antonicci et al. \cite{001}. 

As far as the response to ionizing particles is concerned, bubble nucleation requires that the locally deposited energy of a traversing particle exceeds a defined critical energy necessitated to create a vapour bubble of critical radius $\xRc$, i.e., ${\fonttherm{E_{dep}\geq E_c}}$, and that the stopping power of the particle is large enough to release this energy as heat over a critical deposition length ${\fonttherm{L_{c}}}$ such to be effective to produce a stable vapour bubble of radius $\xRc$, the subsequent growth of the newborn bubble being ensured by the energy supplied by the surrounding liquid, as described by Archambault et al. \cite{06}, which means
\eqb
\fonttherm{E_{dep}=\int^{L_c}_{0} S(E)~ dx \geq E_c}, \label{eq:2}
\eqe
where $S(E)= -dE/dx$ is the stopping power of the ionizing particle of energy $E$, whose values can be calculated using the SRIM package \cite{6}. 

The critical energy $\fonttherm{E_{c}}$ is given by the sum of the energy required to vaporize the mass of liquid involved in the phase change and the energy required to form the vapour bubble surface
 \eqb
\xEc=\fb 4 3\pi\xRc^{3}\xrhoV\xlam +4\pi\xRc^{2}\ (\xsig -\xTL\fb{\d\xsig}{\d\xT} \ ), \label{eq:3}
\eqe
where $\xrhoV$ is the mass density of the saturated vapour and $\xlam$ is the latent heat of vaporization, both evaluated at the vapour temperature $\xTV$, while $\xsig$ is the surface tension of the liquid calculated at the liquid temperature $\xTL$. 

The critical radius $\xRc$, consistent with the condition of mechanical equilibrium between the surface tension and the pressure difference at the bubble surface, can be calculated as the radius of the vapour bubble corresponding to the maximum of the free enthalpy variation associated with the phase change
\eqb
\xRc=\fb{2\xsig }{\xrhoV\xlam \fb{\xTL-\xTV}{\xTV}}. \label{eq:4}
\eqe

In the present investigation, the detector has been operated at a stable nominal liquid temperature $\fonttherm{T_L\mathrm{=25~^oC}}$ with a maximum liquid-vapour temperature difference around $\mathrm{5~^oC}$, which corresponds to a maximum liquid underpressure of about 1 bar. 
Accordingly, the superheat parameter $\fonttherm{SH}$ ranges approximately between 0.05 and 0.1, while the related critical energy $\fonttherm{E_c}$ spans from about 100 keV to nearly 500 keV and the critical radius $\fonttherm{R_c}$ from 80 nm to 140 nm.

Indeed, as thoroughly discussed in a recent study performed by Bruno et al. \cite{006}, a number of equations are available in the literature for the calculation of $\xRc$ and $\xEc$, see e.g. Bugg \cite{011} and Tenner \cite{0111}, to name a few. 
However, in the critical energy range of interest for the current study the use of equations (\ref{eq:3}) and (\ref{eq:4}) or alternative equations proposed by other authors is essentially equivalent, leading to values of $\xEc$ well in agreement within 5 percent. 
Additionally, it seems worth pointing out that, due to the relatively small temperature differences imposed between the target liquid and its vapour, the critical radius $\xRc$ can be reasonably assumed to be proportional to 1/$SH$ --- see equations (\ref{eq:1}) and (\ref{eq:4}) --- which implies that, since the volume term in the critical energy equation is largely predominant with respect to the surface term, the critical energy $\xEc$ can be regarded to be proportional to $R_c\mathrm{^3}$, and then to 1/$SH^{3}$. 
%Actually, in the present investigation range the distribution of $\xEc$ versus $\fonttherm{SH}$  is accurately interpolated by the following equation:
%\eqb
%\xEc~\mathrm{[keV]}= \frac{0,0822}{\xSH^{2.95}}  \label{eq:5}
%\eqe

On the other hand, since $\xRc$ is the natural length scale of the process, the critical deposition length ${\fonttherm{L_{c}}}$ can be expressed in units of the critical diameter of the vapour bubble
\eqb
L_c = k_D \cdot (2R_c) \label{eq:6}
\eqe
where ${\fonttherm{k_{D}}}$ represents the so-called nucleation parameter, whose value, as well as its possible dependence on ${\fonttherm{SH}}$ or ${\fonttherm{E_{c}}}$, are not known a priori, 
but have to be estimated by comparing the experimental counting rates at several operating conditions with the corresponding predictions of Monte Carlo simulations combined with the application of the nucleation model. 
Indeed, different possible values of the nucleation parameter are reported in the literature ranging from 1 to 10 or more, yet, a value of the order of 3, regardless of the nature of the sensitive liquid, seems to find possible theoretical justifications, as e.g. discussed by Norman and Spiegel \cite{0088}, and Bell \cite{0099}. 
Moreover, according to Archambault et al. \cite{06}, the value of ${\fonttherm{k_{D}}}$ should be expected to increase as the critical energy is increased. 
\section{ Background measurements and detector response to\\ $\alpha$-decay events}
\label{sec:3}
\begin{figure} % figuur 1
\begin{minipage}{\columnwidth}
\centering
\vspace{-0.3cm}
\hspace*{-0.5cm} 
\resizebox{0.8\textwidth}{!}{%
  \includegraphics{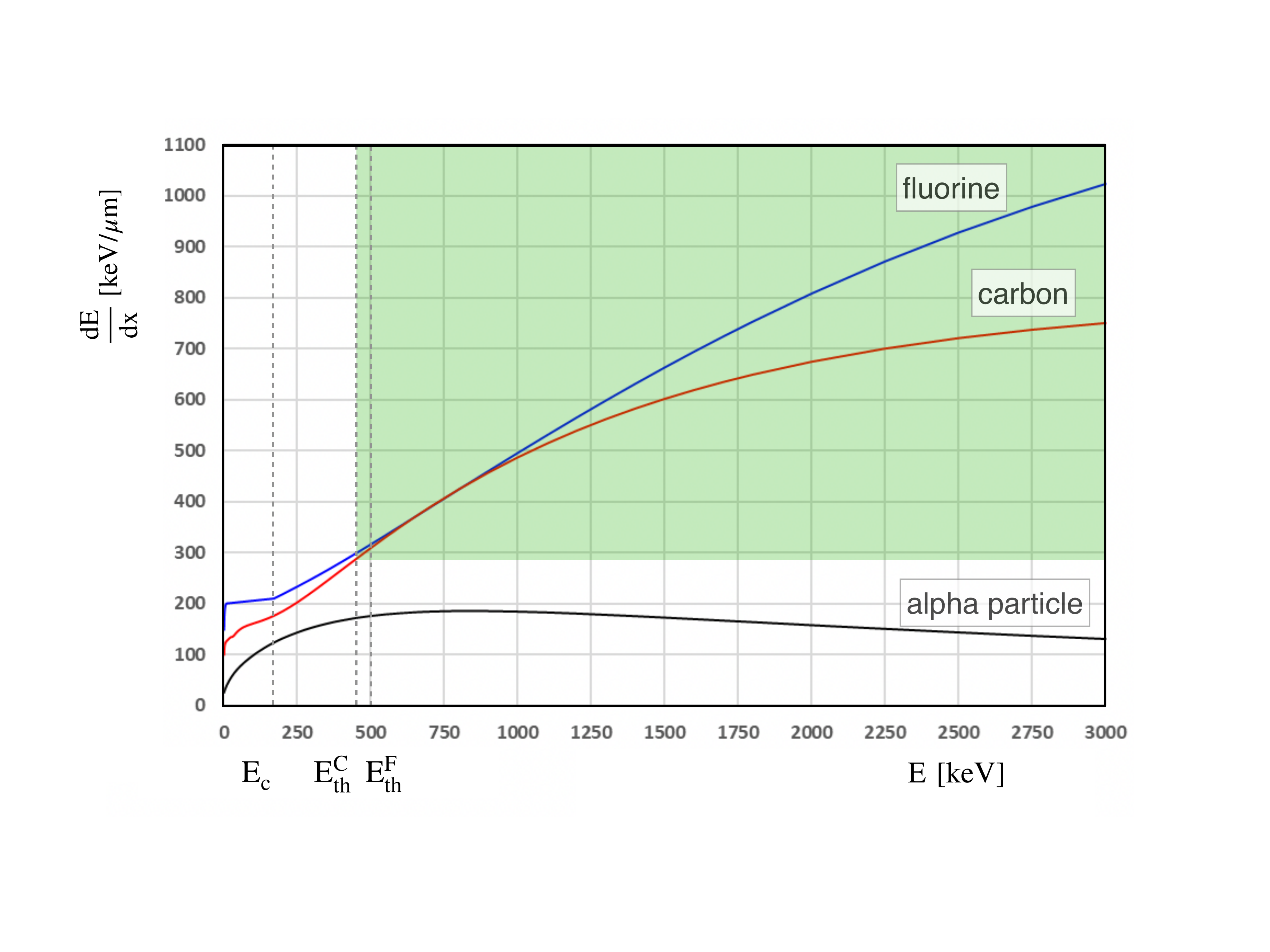}}
%\framebox[\columnwidth][c]{\raisebox{0pt}[20mm][20mm]\includegraphics{Fig_1bw.pdf}}
\end{minipage}
\vspace{-1.cm}
\caption{Stopping power in $\mathrm{C_3F_8}$ at $\mathrm{25^oC}$ for carbon, fluorine and alpha particles (from SRIM package). The region of the parameter space in which nucleation is permitted is highlighted.}
\label{dEdx}
\end{figure}
The main strength of bubble-chamber detectors using superheated liquids is that they can be operated in such thermodynamic conditions to make them insensitive to electron recoils and minimum ionizing particles, which is the case of the aforementioned operating conditions of the MOSCAB detector. 
This means that the only remaining sources available for bubble nucleation are neutrons and $\alpha-$decays occurring inside the liquid. 
However, it must be pointed out that $\alpha-$particles have a stopping power always lower than that of fluorine and carbon ions, as recently discussed by Ardid et al. \cite{0027}. 
This implies that it is possible to find a region in the parameter space ${(E, dE/dx)}$ where the detector maintains its sensitivity to nuclear recoils of fluorine and carbon ions, becoming at the same time insensitive to $\alpha-$particles, whose stopping power, even at the Bragg peak, is lower than that required for bubble nucleation, as displayed in Figure \ref{dEdx}, in which the stopping powers in $\mathrm{C_3F_8}$ at $\mathrm{25~^oC}$ for fluorine and carbon ions are compared with the stopping power typical for an $\alpha-$particle.

Additionally, it must be considered that, although the daughter nuclei emitted in the $\alpha-$decay process, namely $\mathrm{^{218}Po}$, $\mathrm{^{214}Pb}$ and $\mathrm{^{210}Pb}$ in the case of the $\mathrm{^{222}Rn}$ decay chain, have a remarkable stopping power, much higher than $\mathrm{^{12}C}$ and $\mathrm{^{19}F}$, their kinetic energy is rather limited. 
In fact, the highest energy of a naturally emitted $\alpha-$particle is 8.79 MeV, from the decay of $\mathrm{^{212}Po}$ in the $\mathrm{^{232}Th}$ cascade, which implies a kinetic energy of the corresponding recoiling nucleus lower than 170 keV. 
Considering that the energy released by the recoiling nucleus and some amount of the energy deposited by the $\alpha-$particle along its track can add up, it is expected that at a critical energy around 200 keV the detector starts becoming insensitive to $\alpha-$decay processes.
\begin{figure} % figuur 1
\begin{minipage}{\columnwidth}
\centering
\vspace{-0.5cm}
\hspace*{-0.5cm} 
\resizebox{0.8\textwidth}{!}{%
  \includegraphics{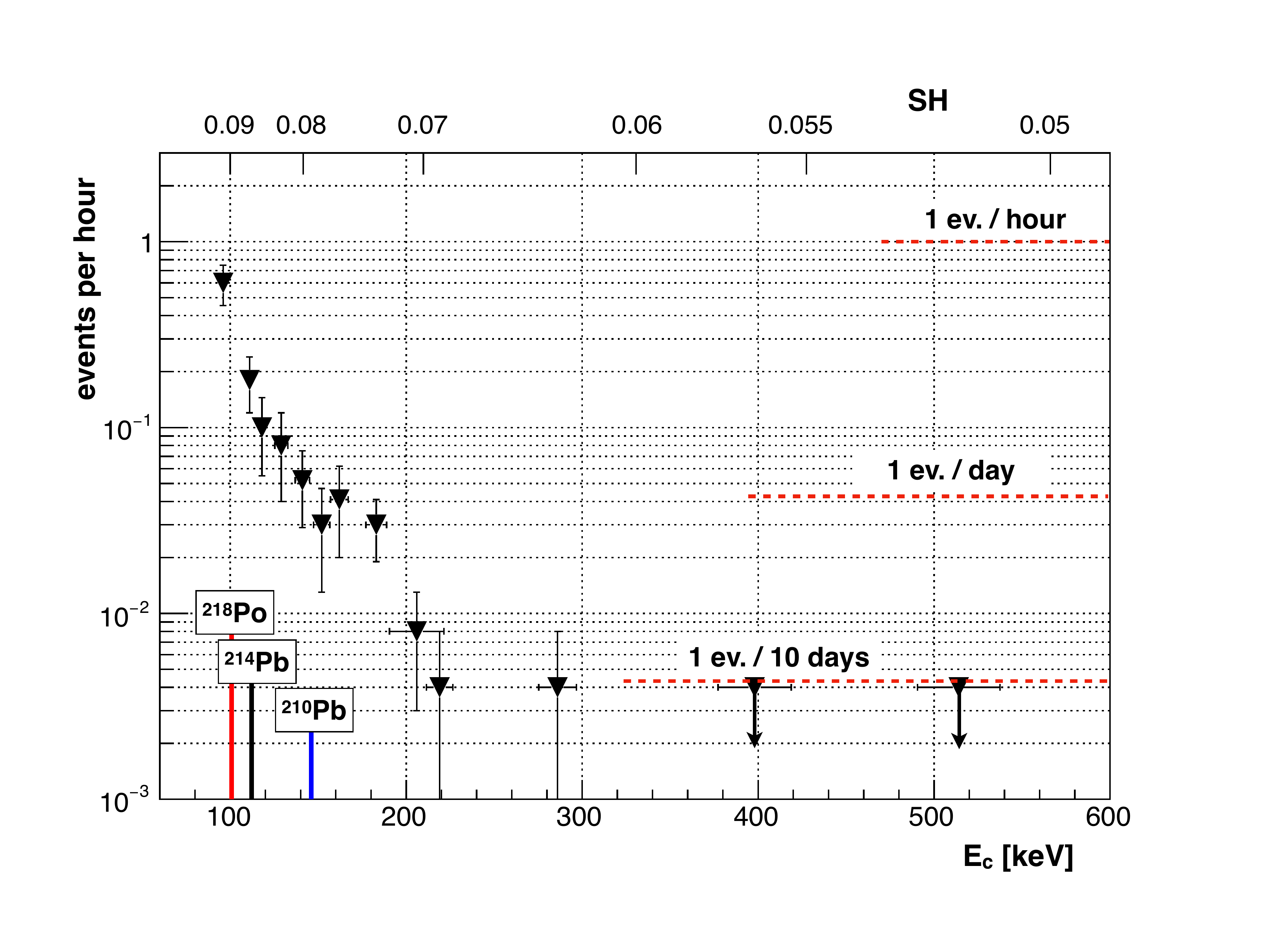}}
%\framebox[\columnwidth][c]{\raisebox{0pt}[20mm][20mm]\includegraphics{Fig_1bw.pdf}}
\end{minipage}
\vspace{-0.5cm}
\caption{Background counting rates of the MOSCAB detector equipped with the 18L vessel filled with 13L of $\mathrm{C_3F_8}$ plotted versus both the critical energy $\xEc$ and the superheat parameter $\xSH$.}
\label{Bk}
\end{figure}
A detailed study of this behaviour has been conducted with the MOSCAB detector located in Hall C of LNGS underground laboratories. 
The detector was equipped with both the 2L and the 18L vessels, filled with 1.2L and 13L of $\mathrm{C_3F_8}$, respectively, and, as said earlier, operated in thermodynamic conditions tuned to obtain a superheat parameter $\xSH$ ranging roughly between 0.05 and 0.1, which corresponds to an approximate critical energy range between 100 keV and 500 keV. 
In particular, a campaign of measurements more than 1800 hours long has been performed from February 2020 to February 2021 to characterize the background of the detector equipped with the 18L vessel, whose results are shown in Figure \ref{Bk}, confirming the results previously obtained using the 2L vessel. 
It is apparent that the background counting rate dramatically decreases as $\xEc$ increases from 100 keV to 200 keV, i.e., as $\xSH$ decreases from 0.09 to about 0.07, reaching values lower than one event per ten days, full in line with the neutron flux at the Gran Sasso underground Laboratory, which confirms that the main intrinsic background source of the detector is represented by $\alpha-$decays mostly due to the $\mathrm{^{222}Rn}$ chain and that when the detector is operated at $\mathrm{SH \leq 0.07}$ its internal background rate
%\textcolor{red}{completely disappears}. 
becomes lower than or at least of the same magnitude of the extremely low environmental neutron flux.

\section{Detector response to fast neutrons } 
\label{sec:4}
\subsection{Data taking}

To measure the detector sensitivity, the bottom of the MOSCAB bubble chamber includes a vertical sleeve allowing to locate a calibration source just below the quartz vessel to minimize the amount of water between the source and the target liquid. 
Therefore, the response to neutrons of the detector operated in the same thermodynamic conditions as those indicated before was measured using a $\mathrm{^{241}}$AmBe neutron source, with an integral neutron strength of nearly 1 n/s. 
\begin{figure} % figuur 1
\begin{minipage}{\columnwidth}
\centering
\vspace{-0.5cm}
\hspace*{-0.5cm} 
\resizebox{0.8\textwidth}{!}{%
  \includegraphics{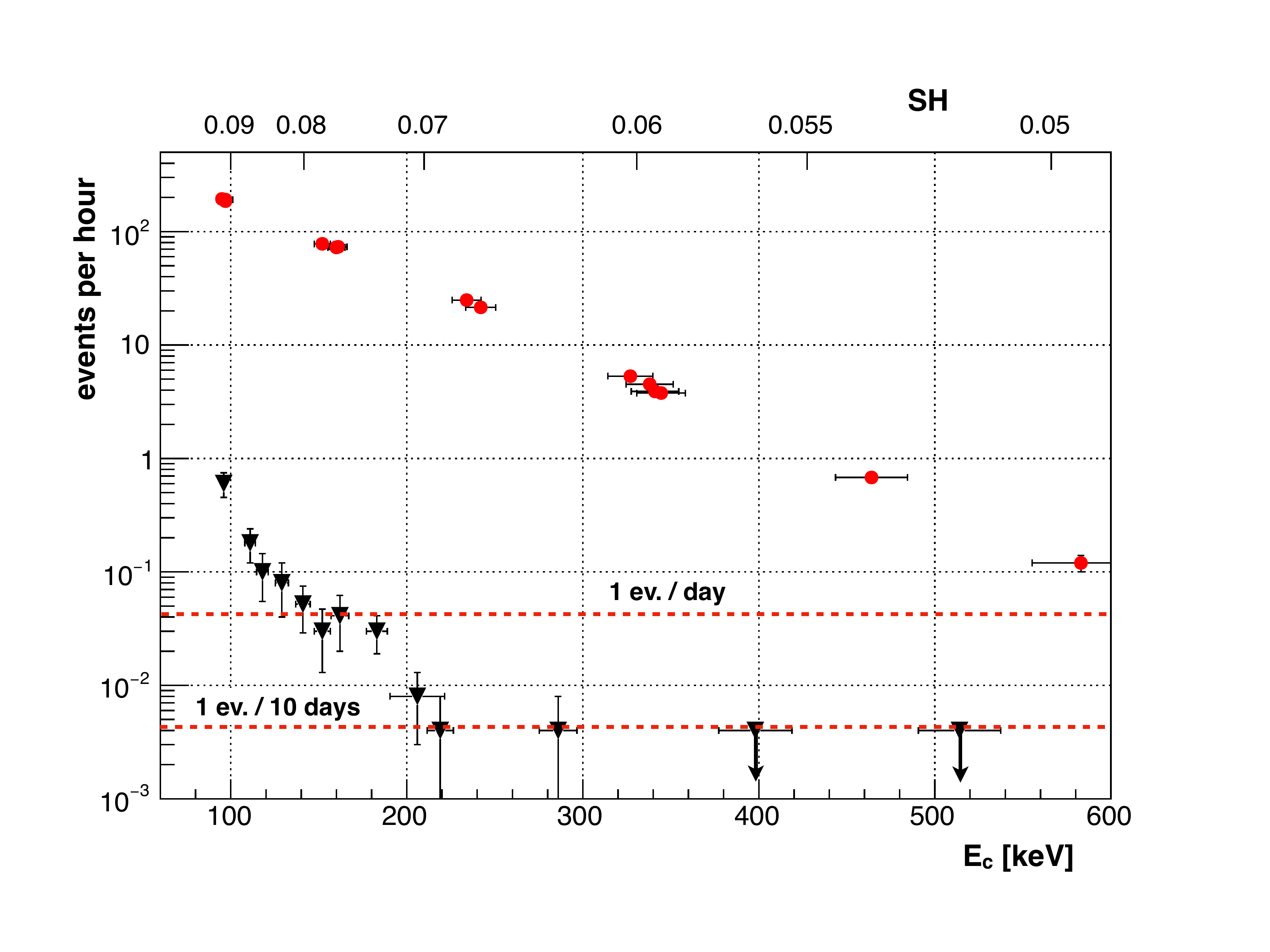}}
%\framebox[\columnwidth][c]{\raisebox{0pt}[20mm][20mm]\includegraphics{Fig_1bw.pdf}}
\end{minipage}
\vspace{-0.5cm}
\caption{Event rates of the MOSCAB detector equipped with the 18L vessel, plotted versus both $\xEc$ and $\xSH$. Full down triangles (black) represent the background counting rate, while the red dots represent the event rate recorded using the $^{241}$AmBe n-source.}
\label{Bk+n1}
\end{figure}
The measurements of the detector response to neutron-induced ion recoils have been performed from December 2019 to April 2021, during more than 2000 hours of data taking. 
The 
%$\mathrm{^{252}Cf}$ 
$^{241}$AmBe source has been used with the detector equipped with the 2L and the 18L vessels. 
%while the $^{241}$AmBe source has been used only with the detector equipped with the 18L vessel. 
The results related to the 18L vessel filled with 13L of $\mathrm{C_3F_8}$ are reported in Figure \ref{Bk+n1}, where the distributions of the recorded event rate are plotted versus the critical energy $\xEc$ and the superheat parameter $\xSH$. 
In the same figure, the background counting rates already shown in Figure \ref{Bk} are also displayed for comparison purposes.  
%\textcolor{red}{
The horizontal error bars represent the standard deviations of $\xEc$ or $\xSH$, which is mainly due to the instrumental uncertainties and hence constant with respect to $\xSH$, whereas the vertical error bars represent the uncertainty of the mean counting rate.
%} 

It is apparent that, even for superheat conditions such that the background counting rate is negligible, i.e., lower than one event per 10 days, the detector maintains its sensitivity to neutron-induced ion recoils.\\

\subsection{Detector response simulation}

The modelling of the detector response is based on the bubble nucleation process described earlier and summarized by equation (\ref{eq:6}), which defines the nucleation parameter $k_{D}$, whose values have been determined by the comparison between each experimental bubble rate detected using the mentioned 
%$\mathrm{^{252}Cf}$ and 
$^{241}$AmBe neutron source located beneath the 2L vessel or the 18L vessel and the numerical results of the Monte Carlo simulations and post-processing carried out in the same thermodynamic conditions.
The effects of the neutron transport and elastic and inelastic interactions with the fluorine and carbon nuclei of the sensitive liquid are evaluated by the way of the MCNP 6.2 code \cite{009} and the SPECTRA-PKA code \cite{010}.
% producing a double-check estimation of the energy distribution of the recoiled ions by the post-processing of the auxiliary file PTRAC, which stores the main information relevant to the neutron interactions occurring inside the target liquid, as well as by implementing the SPECTRA-PKA code \cite{010} using the average neutron flux inside the liquid volume computed by MCNP as input data.
The energy distributions of the recoiled ions, in conjunction with the stopping power data for $\mathrm{^{19}F}$ and $\mathrm{^{12}C}$ in $\mathrm{C_3F_8}$ at temperature $T_L$ calculated by the way of the SRIM code, are then used to determine the critical deposition length $L_c$ along which, in the continuous slowing-down approximation, a recoiled ion deposits an average energy $E_{dep}$ at least equal to the critical energy $E_c$, such that the simulated and experimental counting rates are the same.

Accordingly, denoting as $(d \dot{N}_r/dE)_C$ and $(d \dot{N}_r/dE)_F$ the rate of carbon and fluorine recoiled ions per unit energy, respectively,
%the recoil energy distribution of the i-th ion as $\mathnormal{\(\frac{dr}{dE_r}\)_i}$, 
a pair of threshold recoil energies $E_{th}^C$ and $E_{th}^F$ can be determined for each value of $\mathnormal{E_c}$ such that the nucleation rate $\dot{N}_{ev}(E_c,k_D)$ is
%the rate of bubble nucleation is
%\eqb
%\mathnormal{R_{}(E_c, k_D)= \sum\limits_{i\in\{F,C\}} \int^{\infty}_{E_c}H~[L_c - L_{dep}] \cdot \(\frac{dr}{dE_r}\)_i~dE_r } \label{eq:7}
%\eqe
%in which $\mathnormal{H~[\cdot]}$ is the Heaviside step function: 
%\eqb
%\mathnormal{H~[L_c - L_{dep}]  = H~\[k_D\mathrm{2}R_c - \int^{E_0}_{E_0-E_c}\(\frac{\mathrm{1}}{S(E)}\)_i dE\]}  \label{eq:8}
%\eqe
%and $\mathnormal{L_{dep}}$ is the average deposition length required to any ion aving an initial energy $E_0$ to release exactly the critical energy $\mathnormal{E_c}$, calculated by integrating the reciprocal of the stopping power over energy.\\
%
%Hence, 
\eqb
\dot{N}_{ev}(E_c, k_D)= \int^{\infty}_{E_{th}^C} \left( \frac{d \dot{N}_r}{dE}\right)_C~dE + \int^{\infty}_{E_{th}^F} \left( \frac{d \dot{N}_r}{dE}\right)_F~dE
. \label{eq:9}
%   +\int^{\infty}_{E_{th}^F} {(\frac{d \dot{N_r}}{dE})_F~dE }
\eqe
Thus, once the numerical counting rate is imposed to be the same as the experimental one and the critical deposition length $L_c$ is assumed to be the same for any ionizing particle, the value of $k_D$ can be calculated by integrating the reciprocal of the stopping power for either $^{12}$C or $^{19}$F
\begin{equation}
\mathnormal{k_D(E_c) = \frac{\mathrm{1}}{\mathrm{2}R_c(E_c)} \int^{E_{th}^F}_{E_{th}^F-E_c} \frac{{dE}}{S_F(E)} = \frac{\mathrm{1}}{\mathrm{2}R_c(E_c)} \int^{E_{th}^C}_{E_{th}^C-E_c}\frac{{dE}}{S_C(E)}}.  \label{eq:10}
\eqe
\begin{figure} % figuur 1
\begin{minipage}{\columnwidth}
\centering
\vspace{-1.5cm}
\hspace*{-0.5cm} 
\resizebox{0.8\textwidth}{!}{%
  \includegraphics{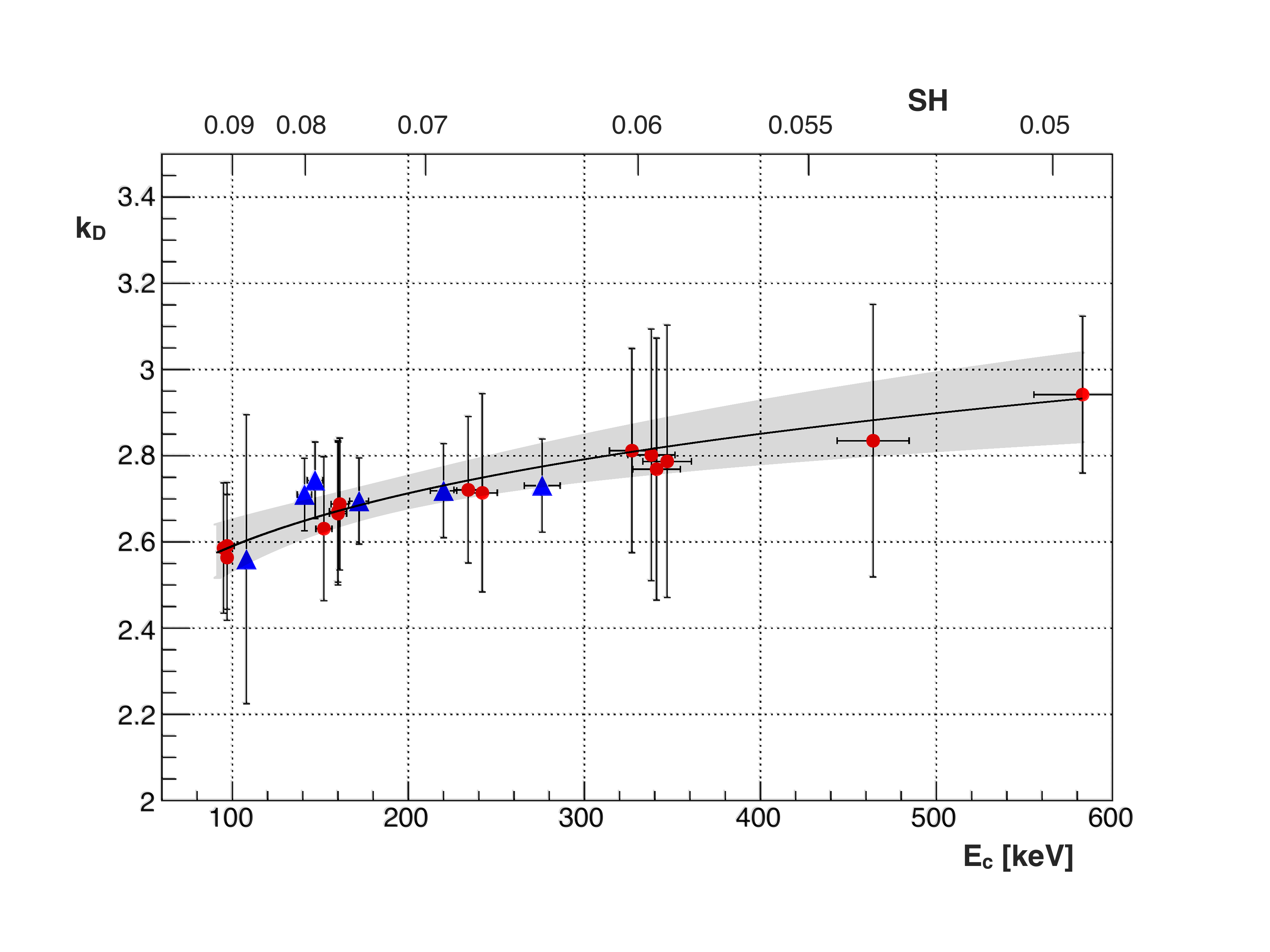}}
%\framebox[\columnwidth][c]{\raisebox{0pt}[20mm][20mm]\includegraphics{Fig_1bw.pdf}}
\end{minipage}
\vspace{-0.5cm}
\caption{The nucleation parameter $k_D$ as a function of the critical energy $\xEc$ and the superheat parameter $\xSH$. The red dots and blue triangles are obtained by the measurements with the 18L vessel and the 2L vessel respectively. The best fit curve is shown with the 1$\sigma$ region shaded in grey.}
\label{k}
\end{figure}

The distribution of the values obtained for the nucleation parameter $k_D$ is displayed in Figure \ref{k} versus the critical energy $E_c$  and the superheat parameter $\xSH$, showing that the critical deposition length is practically independent of the volume of the sensitive liquid, which means that $k_D$ can be considered to be a characteristic of the target liquid and its degree of metastability.
% and then that
%, since the effects of the interactions occurring between any recoiled ion and the sensitive liquid depend on the energy of the recoiled ion, which is only related to the energy of the scattering neutron, regardless of the spectrum of the emitting source,
%the nucleation parameter $k_D$ is a characteristic of the superheated fluid. 
%
%
As far as the interpolation curve is concerned, it must be noticed that within the investigated energy range the electronic stopping power for both fluorine and carbon ions is largely dominant with respect to the nuclear counterpart and, since its value is directly proportional to the ion speed, as described by Lindhard \cite{012}, the solution of each integral in equation (\ref{eq:10}) can be assumed to be proportional to $E_c^{1/2}$.
Therefore, taking into account the afore-mentioned proportionality relation $R_c \propto E_c^{1/3}$, the nucleation parameter $k_D$ is expected to be proportional to $E_c^{1/6}$, which can be expressed using the general form:
\eqb
\mathnormal {k_D (E_c) =  a \cdot \sqrt [6] {E_c} + b}.  \label{eq:11}
\eqe
Actually, a satisfactory best fit of the data is obtained when the values of the two empirical constants $a$ and $b$ are assumed to be 0.42 and 1.70, as shown in Figure \ref{k} 
where the 1$\sigma$ region is shaded in grey. 
%calculated in the hypothesis of Gaussian fluctuations of both $E_c$ and $k_D$ equal to 0.23 and 0.54, respectively.
%
\begin{figure} % figuur 1
\begin{minipage}{\columnwidth}
\centering
\vspace{-0.5cm}
\hspace*{-0.5cm} 
\resizebox{0.8\textwidth}{!}{%
  \includegraphics{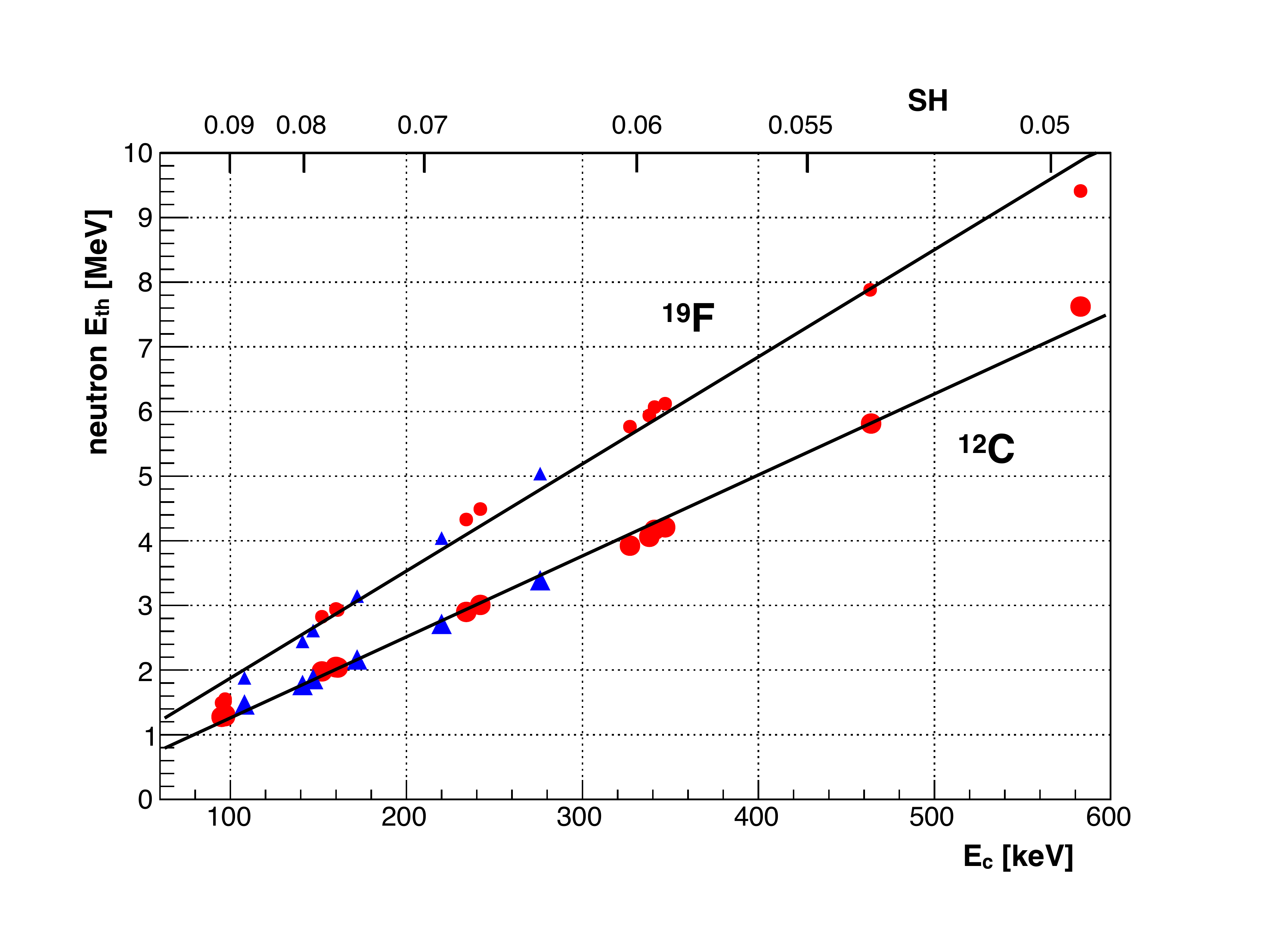}}
%\framebox[\columnwidth][c]{\raisebox{0pt}[20mm][20mm]\includegraphics{Fig_1bw.pdf}}
\end{minipage}
\vspace{-0.5cm}
\caption{The neutron energy threshold as a function of $\xEc$ and $SH$.
The red dots and blue triangles are obtained by the measurements with the 18L vessel and the 2L vessel respectively. Small for F nuclei and large for C nuclei.}
\label{Eth}
\end{figure}

The procedure described above allows to determine a direct relationship between the metastability degree at which the equipment is operated, identified by the reduced superheat parameter $SH$,
and the neutron energy threshold $E_{th}^n$.
In fact, the recoil energy thresholds $E_{th}^C$ and $E_{th}^F$ establish the minimum amount of kinetic energy needed to the recoiled ion to produce a bubble nucleation and then, taking into account the kinematic factor, the corresponding neutron energy threshold $E_{th}^n$. 
Indeed, in the entire investigated energy range such a neutron energy threshold is determined by carbon, as displayed in Figure \ref{Eth}, where the distributions of $E_{th}^n$, due to recoiling $\mathrm{^{12}C}$ and $\mathrm{^{19}F}$ nuclei, are plotted versus both $E_c$ and the $\xSH$.
According to the data, the relationships existing between the neutron threshold $E_{th}^n$ and both the critical energy and the superheat parameter are well approximated by the following equations
\begin{eqnarray}
\mathnormal {E_{th}^n (E_c)} = (12.5\pm0.2) E_c + (7.5\pm70)  \label{eq:12}\\
\mathnormal {E_{th}^n (SH) =  \mathrm{\frac{(1.036 \pm 0.016)}{\mathnormal{SH}^{2.95}} + (7.5\pm70)} } \label{eq:13}
\end{eqnarray}

\section{Efficiency curves }
\label{sec:5}

\begin{figure}[h!] % figuur 1
\begin{minipage}{\columnwidth}
\centering
\vspace{-0.2cm}
\hspace*{-0.5cm} 
\resizebox{0.8\textwidth}{!}{%
  \includegraphics{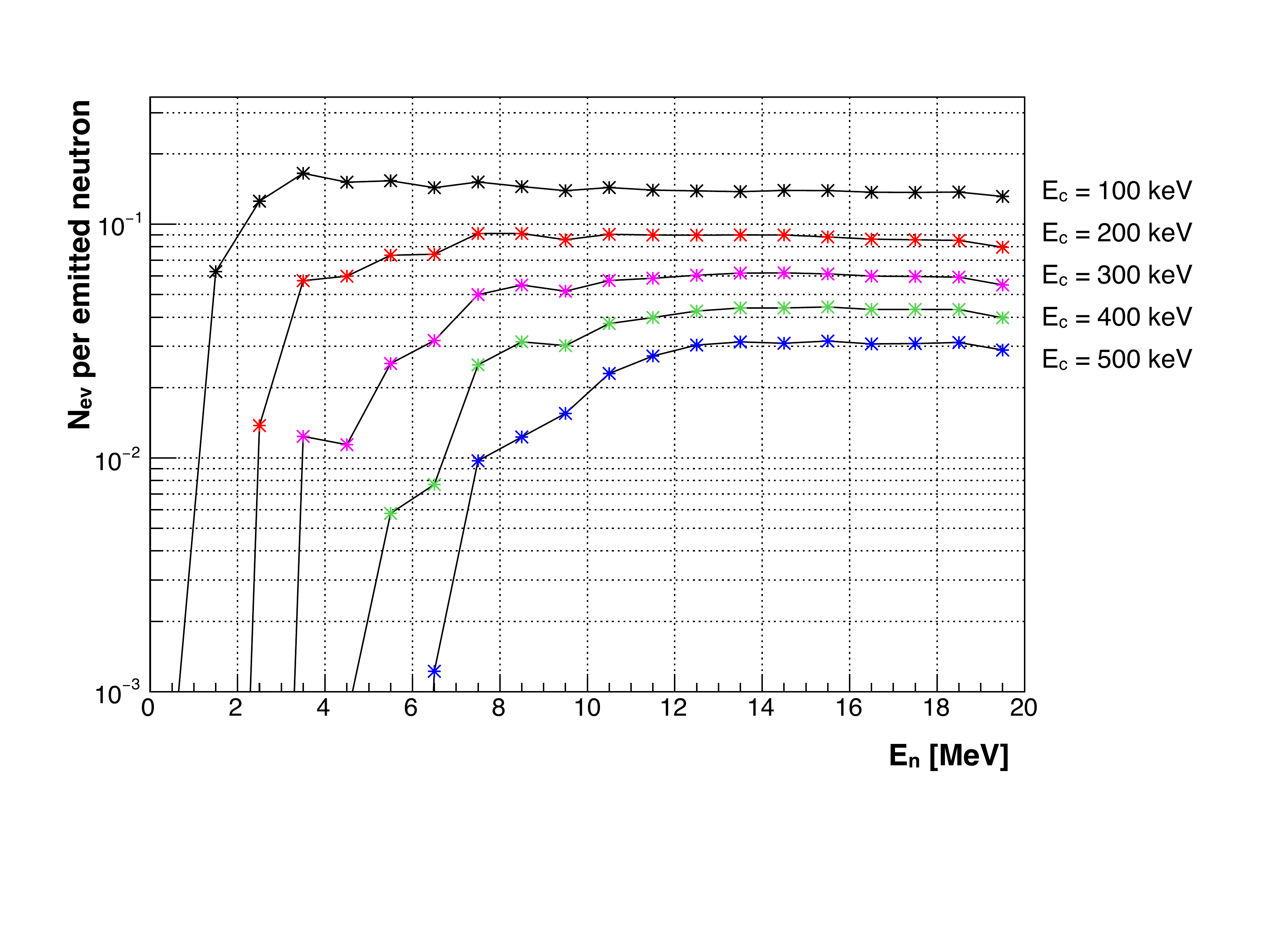}}
%\framebox[\columnwidth][c]{\raisebox{0pt}[20mm][20mm]\includegraphics{Fig_1bw.pdf}}
\end{minipage}
\vspace{-1.cm}
\caption{Efficiency functions of the MOSCAB detector versus the neutron energy. The efficiencies curves are calculated for an isotropic point-like source located at the bottom of the 18L vessel filled with 13L of $\mathrm{C_3 F_8}$ and different value of the critical energy $\xEc$.}
\label{resp1}
\end{figure}
The bubble nucleation model based on equations (\ref {eq:2})-(\ref {eq:6}) and (\ref {eq:11}) exhaustively describes the behaviour of MOSCAB detector, at least in the critical energy range 
$\mathrm{100~keV\leq\xEc}$ 
$\mathrm{\xEc \leq 500~keV}$, 
thus allowing to generate, via MCNP simulations and subsequent data processing, sets of response functions using mono-energetic neutron sources. 

The detection efficiency curves computed for an isotropic point-like source located just below the bottom of the 18L vessel filled with 13L of $\mathrm{C_3F_8}$ are displayed in Figure \ref{resp1}, 
where the distributions of the expected number of nucleation events per emitted neutron are plotted against the neutron energy, using the critical energy $\xEc$ as a parameter. 
Similar response functions have also been obtained using the 2L vessel filled with 1.2L of $\mathrm{C_3F_8}$. 
Notice that, due to the location of the source and its isotropic emission, the maximum reachable detection efficiency is of about 35\% for the 18L vessel, and about 20\% for the 2L vessel due to the different geometry of the two jars.

%the maximum reachable detection efficiency is of about 35\% for the 18L vessel, and about 20\% for the 2L vessel.\\
%
The obtained efficiency curves have then been employed to determine the rates of nucleation events consequent to the emission by the $^{241}$AmBe neutron source used in the experimental campaign. 
The comparison between the simulated counting rates and the experimental data is reported in Figure \ref{rates} for the 2L and the 18L vessels. 
The excellent agreement proves the reliability of the developed nucleation model, which proves to be effective regardless of the amount of the target liquid and the geometry of the vessel.\\
\begin{figure} % figuur 1
\begin{minipage}{\columnwidth}
\centering
\vspace{-1.cm}
\hspace*{-0.5cm} 
\resizebox{0.8\textwidth}{!}{%
  \includegraphics{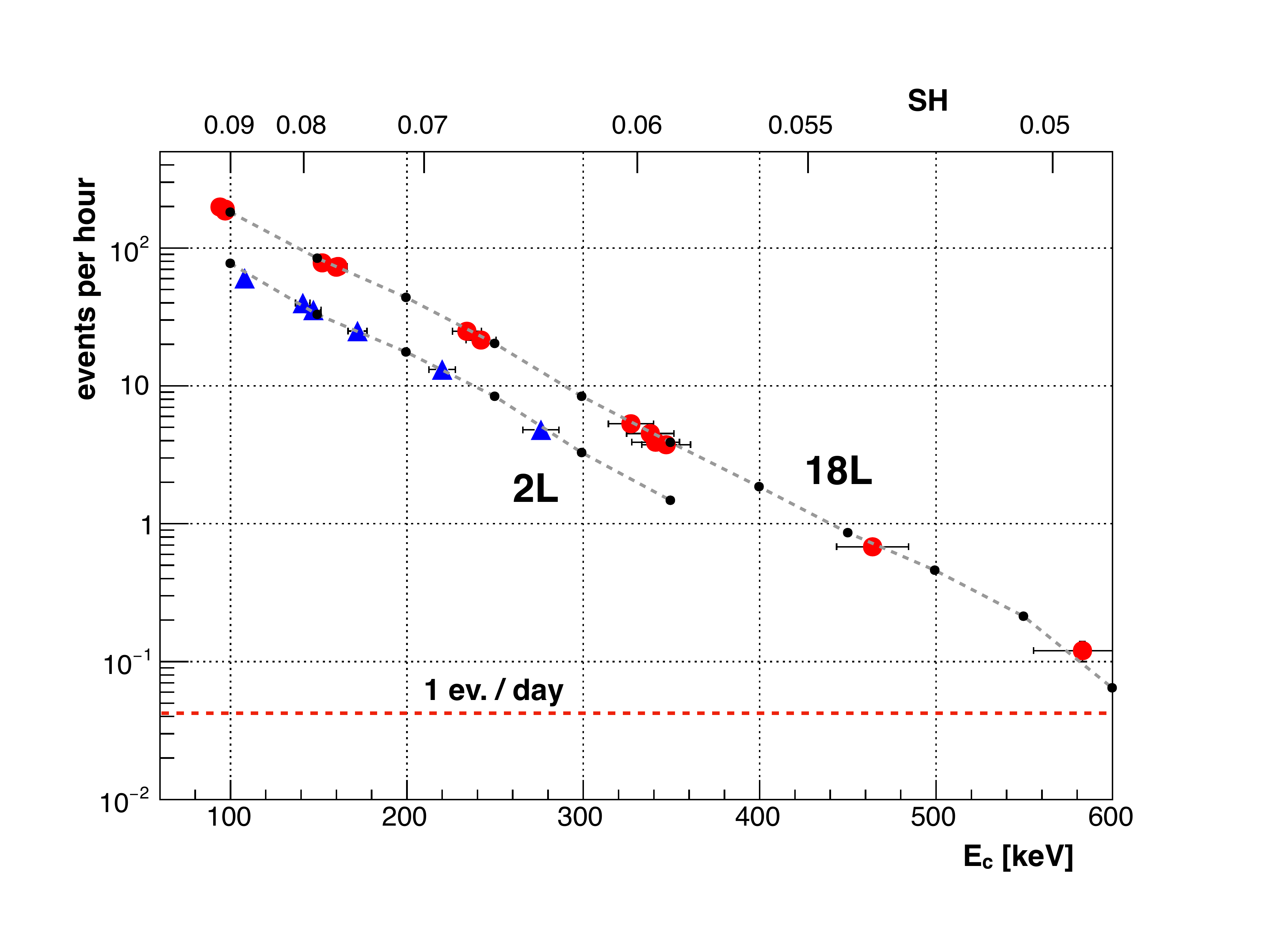}}
%\framebox[\columnwidth][c]{\raisebox{0pt}[20mm][20mm]\includegraphics{Fig_1bw.pdf}}
\end{minipage}
\vspace{-0.5cm}
\caption{Event rate response of the MOSCAB detector to the $^{241}$AmBe neutron source equipped with the 18L and the 2L vessels filled with $\mathrm{C_3F_8}$, plotted versus both $\xEc$ and $\xSH$. 
The red dots and blue triangles represent the counting rate detected inside the 18L and the 2L vessels respectively.
The black dots represent the corresponding results from the Monte Carlo simulations.}
\label{rates}
\end{figure}

Thus, using this bubble-chamber detector, placed in a clean environment, i.e., an underground laboratory, it is possible to characterize a neutron source and, if the energy spectrum is known, determine its activity with very high sensitivity in a reasonable time.
%having a well-defined energy spectrum.
In fact, if the rate of neutrons emitted by the source per unit energy is written as
\eqb
\frac{d \dot{N}_n}{dE} =  A \cdot g_n(E)  \label{eq:14}
\eqe
where A is the source activity and $g_n(E)$ is the normalized spectrum of the neutron source, then the 
%with the condition:
%\eqb
%\mathnormal {\int{g(E_n)dE_n} = \mathrm{1}}  \label{eq:16}
%\eqe
%and $G(E_c,E_n) $ is the response function of Figure \ref{resp1} expressed as bubbles per source neutron, 
nucleation rate detected when the critical energy is set to $\xEc$ can be expressed as
 \eqb
\dot{N}_{ev}(E_c) = \int^{\infty}_0{G(E_c,E) \cdot A \cdot g_n(E)dE}   \label{eq:15}
\eqe
in which $G(E_c,E) $ is the efficiency function describing the response of the detector to neutrons emitted by a point-like source located in the source-housing of the equipment.
Accordingly, the minimum detectable source activity 
%at 90\% c.l. 
can be evaluated as: 
\eqb
A_{min}(E_c)=\frac{\dot{N}_{s}(E_c)}{\int^{\infty}_0{G(E_c,E) \cdot g_n(E)dE}} \label{eq:16}
\eqe
in which $\dot{N}_{s}$ represents the 99\% c.l. upper limit of the background counting rate, or the lower discernible counting rate due to a neutron source.

If the external background conditions are the same as those of the LNGS underground laboratory, the results achieved for an isotropic point-like neutron source located inside the detector source-housing, for example a $\mathrm{^{252}}$Cf source, are listed in Table \ref{sens1} for different values of the critical energy $\xEc$. 
Thus, a $\mathrm{^{252}}$Cf source with an activity of $\mathrm{1\cdot10^{-3}}$ neutrons per second could be disentangled from the background at 99\% c.l. by a measurement having a duration of the order of ten days.
%As an example, the results achieved for an isotropic $\mathrm{^{252}}$Cf neutron source located inside the detector source-housing are listed in Table \ref{sens1} for different values of the critical energy $\xEc$. 
%Based on the results obtained, a $\mathrm{^{252}}$Cf neutrons source activity of few neutrons per hour can be obtained, in the LNGS external background conditions and with an uncertainty of 10\%, by a measurement having a duration of a few tens of days.
%be measured during some day of measurement at 90\% c.l.
%For the $\mathrm{^{252}}$Cf neutron energy spectrum, the Watt spectrum with the two parameters $T_w$ = 1.18 MeV and $E_w$ = 0.36 MeV \cite{008} has been used.
%
Of course, should both the activity and the energy spectrum of the source be unknown, the coupling of the procedure discussed above with an unfolding technique can effectively be used to fully characterize the neutron source emission. 

\begin{table}[h!]
\vspace{-0.cm} 
\caption{Minimum detectable source activity, $\mathrm{A_{min}}$, for a $\mathrm{^{252}}$Cf source placed inside the detector source-housing, if the external background conditions are the same as those of the LNGS underground laboratory.}
% and minimum detectable neutron flux $\mathrm{\Phi_{min}}$, at 99\% c.l.}

\label{sens1}
%\small
\vspace{.3cm}
%\begin{tabular*}{\columnwidth}{@{\extracolsep{\fill}}|c|c|c|c|@{}}
\begin{tabular*}{\columnwidth}{@{\extracolsep{\fill}}llll@{}}
\hline
$\xEc~[keV]$ & $\mathrm{\dot{N}_{s}~[{ev}\cdot sec^{-1}]}$ & $\int^{\infty}_0{G(E_c,E) g_n(E)dE}$ &$\mathrm{A_{min}~[n\cdot sec^{-1}]}$\\
\hline
100  & $~~~ \mathrm{1.91 \cdot 10^{-4}}$ & $\mathrm{7.80\cdot 10^{-2}}$ & $\mathrm{2.45\cdot 10^{-3}}$ 
%& $\mathrm{1.15\cdot 10^{-6}}$ 
\\
%\hline
200  &$~~~\mathrm{1.44 \cdot 10^{-5}}$ & $\mathrm{1.78 \cdot 10^{-2}}$ & $\mathrm{8.09\cdot 10^{-4}}$ 
%& $\mathrm{9.82\cdot 10^{-8}}$
\\
%\hline
300  &$~~~ \mathrm{9.00\cdot 10^{-6}}$ & $\mathrm{4.42 \cdot 10^{-3}}$ & $\mathrm{2.04\cdot 10^{-3}}$ 
%& $\mathrm{3.70\cdot 10^{-8}}$
\\
400  &$~~~\mathrm{5.33\cdot 10^{-6}}$ & $\mathrm{9.08 \cdot 10^{-4}}$ & $\mathrm{5.87\cdot 10^{-3}}$ 
%& $\mathrm{4.05\cdot 10^{-8}}$
\\
500  & $~~~\mathrm{5.33\cdot 10^{-6}}$ & $\mathrm{2.64 \cdot 10^{-4}}$ & $\mathrm{2.02\cdot 10^{-2}}$ 
%&
\\
\hline
\end{tabular*}
\end{table}
\begin{figure} % figuur 1
\begin{minipage}{\columnwidth}
\centering
\vspace{-2.5cm}
\hspace*{-0.5cm} 
\resizebox{0.8\textwidth}{!}{%
  \includegraphics{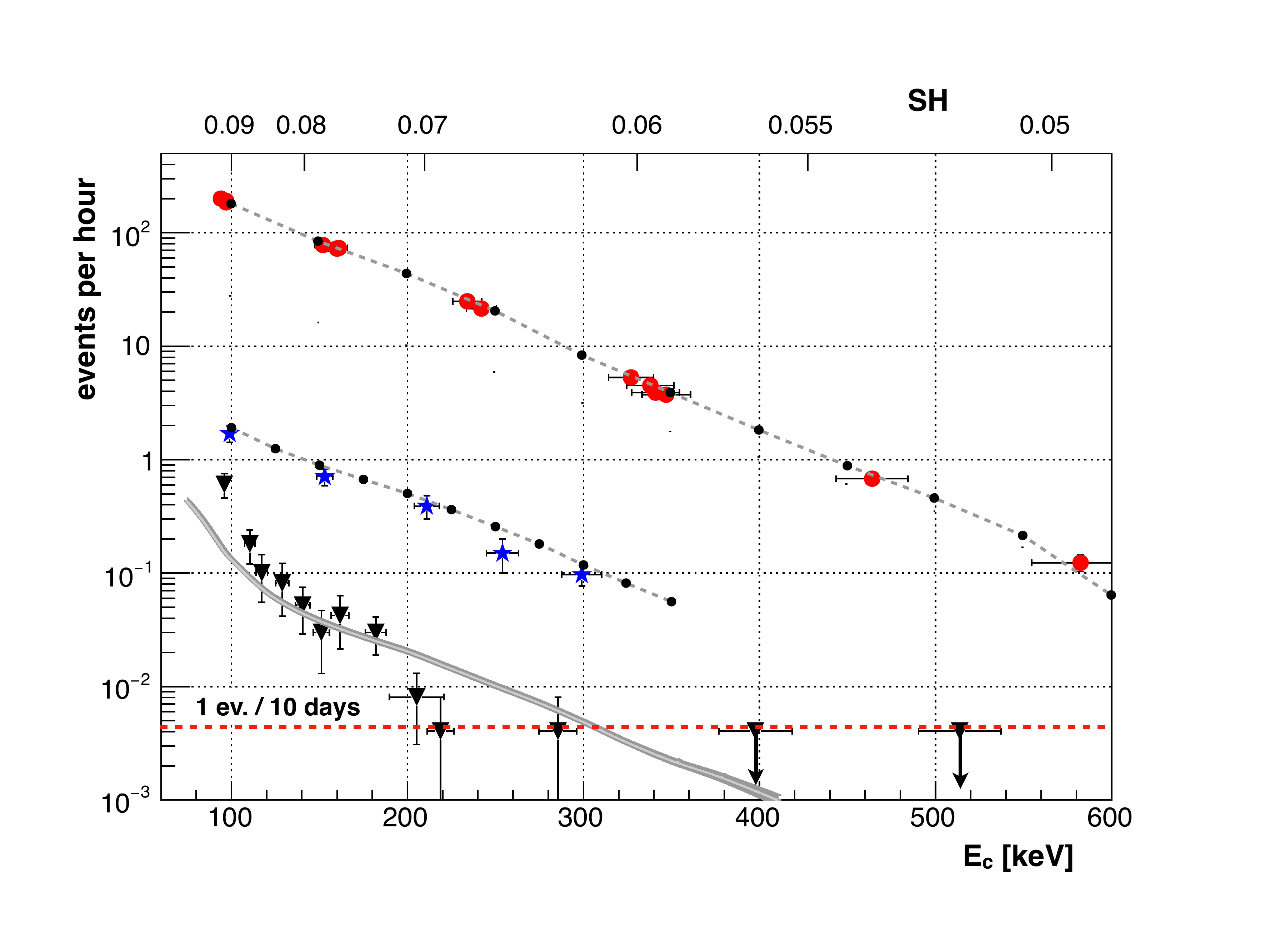}}
%\framebox[\columnwidth][c]{\raisebox{0pt}[20mm][20mm]\includegraphics{Fig_1bw.pdf}}
\end{minipage}
\vspace{-0.5cm}
\caption{Event rates of the MOSCAB detector equipped with the 18L vessels, plotted versus both $\xEc$ and $\xSH$. 
The red dots represent the detector counting rate when the $^{241}$AmBe source is located inside the source-housing, while the blue stars represent the counting when the same source is placed in contact with the outer surface of the detector. 
The black dots reports the corresponding results from the simulations. 
The background counting rate, represented using black down triangles, is also reported and compared with the counting rate expected from the environmental fast neutron flux as calculated by Wulandari et al. \cite{002} for the Hall C of the Gran Sasso Laboratory (gray line).}
\label{fuori}
\end{figure}

The validation of the simulation of the detector behaviour, which the efficiency curves are derived from, has been carried out by changing the experimental conditions, i.e., by moving the $^{241}$AmBe source from the source-housing at the bottom of the quartz vessel to the outside of the detector, in direct contact with its external surface, again using the 18L vessel filled with 13L of $\mathrm{C_3F_8}$.
The new counting rates plotted versus both $\xEc$ and $\xSH$ are displayed in Figure \ref{fuori}, where the data previously obtained with the same neutron source located inside the source-housing of the detector are also reported for comparison, revealing that the presence of the water thermal bath surrounding the quartz vessel containing the target liquid, as well as the stainless-steel boundary wall of the detector, give rise to an attenuation of the neutron flux emitted by the $^{241}$AmBe source of the order of $\mathrm{10^2}$. 
In the same figure, 
the expected rates obtained by the Monte Carlo simulations and subsequent post-processing are also represented using black dots, whose satisfactory degree of agreement with the experimental data makes us confident enough in the reliability of both the nucleation model and the detector simulation procedure. \\
% (335 mm from the bottom, 165 degree clock wise from the bisector of the two cameras).

\section{Detector sensitivity}

\begin{figure}[h!] % figuur 1
\begin{minipage}{\columnwidth}
\centering
\vspace{-0.2cm}
\hspace*{-0.5cm} 
\resizebox{0.8\textwidth}{!}{%
  \includegraphics{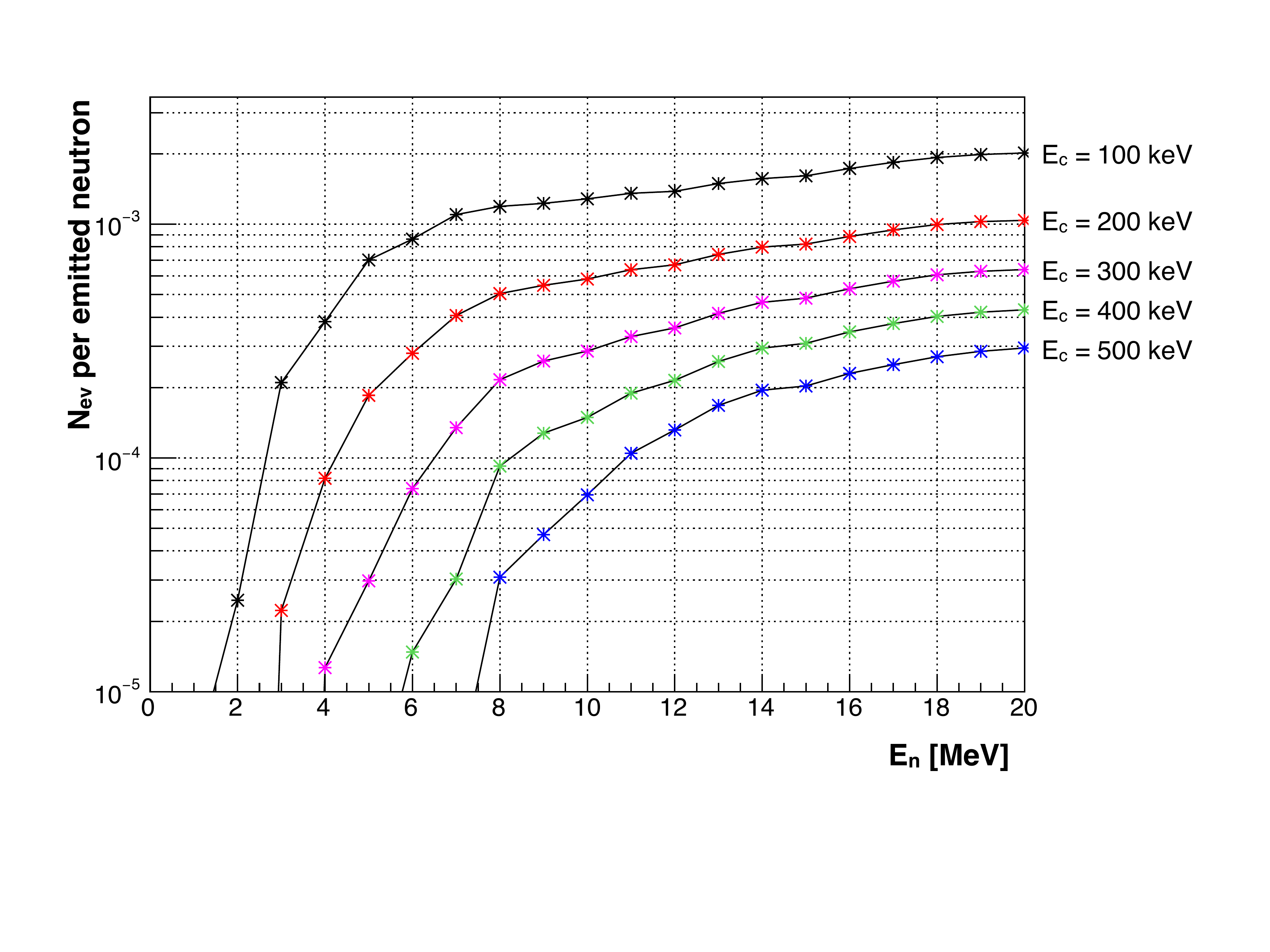}}
%\framebox[\columnwidth][c]{\raisebox{0pt}[20mm][20mm]\includegraphics{Fig_1bw.pdf}}
\end{minipage}
\vspace{-1.cm}
\caption{Efficiency functions of the MOSCAB detector versus the neutron energy, calculated for a spherical (3m diameter) isotropic neutron source surrounding the detector equipped with the 18L vessel filled with 13L of $\mathrm{C_3 F_8}$ and different value of the critical energy $\xEc$.}
\label{resp2}
\end{figure}
The efficiency functions $G'(E_c,E)$ of the MOSCAB detector affected by a diffuse flux of neutrons,
obtained by simulating a neutron emission by a spherical surface having a diameter of 3m inside which the detector is enclosed,
are reported in Figure \ref{resp2}, the change in slope occurring around E~=10~MeV being due to the rise of the n-capture cross sections of the oxygen contained into the water of the thermal bath.
Following the same procedure employed for a neutron source located inside the source-housing, the minimum detectable diffuse neutron flux at 99\% c.l. can be calculated as:
\eqb
\Phi_{min}(E_c)=\frac{\dot{N}_{s}(E_c)}
{S \cdot \int^{\infty}_0{G'(E_c,E) \cdot g_n(E)dE}} \label{eq:17}
\eqe\\
where $S$ is the surface area of the sphere surrounding the detector, from which the incoming simulated neutrons take origin.
%For a neutron source having a constant spectrum up to $E \mathrm{= 20~ MeV}$, the minimum detectable neutron fluxes, at 90\% c.l., are listed in the fifth column of Table \ref{sens1} for different values of the critical energy $\xEc$.\\
%
%The number of neutrons per unit time and unit area emitted by the source can be written, in this case, in the form:
%\eqb
%\frac{d\dot{N}_n}{dE_n}dE_n = { \Phi} \cdot g(E_n)  \label{eq:19}
%\eqe
%In this case, the nucleation rate is:
%\eqb
%\dot{R}(E_c) =S \cdot  \int{G(E_c,E_n) \cdot \Phi \cdot g(E_n)dE_n}   \label{eq:15}
%\eqe
%and the minimum detectable flux, at 90\% c.l.:
%\eqb
%\Phi_{min}(E_c)=\frac{\dot{R}_{s}(E_c)}{ S \cdot \int{G(E_c,E_n) \cdot g(E_n)dE_n}} \label{eq:16}
%\eqe\\
%where $\Phi$ is the neutron flux ($\mathrm{n\cdot cm^{-2}sec^{-1}}$)
%, $g(E_n)$ the normalized energy spectrum 
%and ${S}$ the surface of the sphere surrounding the detector, origin of simulated neutrons. \\
%In the case of a neutron spectrum constant for $E_n \leq 20~ MeV$, the minimum detectable flux $\Phi_{min}$ is shown in Table \ref{sens2} for different values of the critical energy $\xEc$.\\

As an example of application of equation (\ref{eq:17}) the contribution of the environmental background of LNGS to the background counting rate of the detector has been calculated.
A detailed discussion on the low energy component of the neutron flux inside LNGS underground laboratories can be found in the study executed on this topic by Wulandari et al. \cite{002}, according to which the flux is dominated by neutrons produced in the concrete layer, thus implying that the neutron flux is practically independent of the specific location considered. 
Fission and ($\mathrm{\alpha , n}$) reactions contribute to the total production rates but, while the spontaneous fission of $\mathrm{^{238}U}$ mainly produces neutrons having energies below 4 MeV, the ($\mathrm{\alpha , n}$) reactions are substantially responsible for the production of neutrons with higher energies.
Of course, an increase in the water content of the rock or the concrete layer can determine a neutron flux decrease due to the higher hydrogen content and a consequently higher moderation efficiency.
Thus, based on the neutron spectrum calculated by Wulandary and co-workers for the Hall C of LNGS, the efficiency curves displayed in Figure \ref{resp2} are used to determine the nucleation event rate $\mathrm{\dot{N}_{W}}$ expected inside the detector equipped with the 18L vessel filled with 13L of $\mathrm{C_3F_8}$, as listed in Table \ref{sens3}, in which the measured background rate $\mathrm{\dot{N}_{bk}}$ is also reported for comparison.  

It is worth noticing that for $\xEc \leq$ 100 keV the event rate of the detector due to internal $\alpha$-decay events exceeds the environmental background, while for $\xEc \geq $ 300 keV both the calculated and measured event rates, become lower than one event per ten days. 
In the energy region between these two values, the experimental counting rate can be ascribed to be due to the environmental background, as shown in Figure \ref{fuori} and reported in Table \ref{sens3}.
\begin{table}[h!]
\centering 
\caption{ MOSCAB 18L background counting rate as a function of the critical energy, $\xEc$ in comparison with the counting rate expected from the environmental fast neutron flux as calculated by Wulandari et al. \cite{002} for the Hall C of the Gran Sasso Laboratory. }

\label{sens3}
\vspace{.3cm}
\begin{tabular*}{\columnwidth}{@{\extracolsep{\fill}}lll@{}}
\hline
$\xEc~[keV]$ & $\mathrm{\dot{N}_{bk}~[{ev}\cdot h^{-1}]}$ & $\mathrm{\dot{N}_{W}~[{ev}\cdot h^{-1}]}$  \\
\hline
$\mathrm{152 \pm5}$  & $ \mathrm{~(3.0 \pm 1.7 )\cdot 10^{-2}}$ & $\mathrm{3.71 \cdot 10^{-2}}$ 
\\
$\mathrm{162 \pm5}$  & $ \mathrm{~(4.1 \pm 2.1) \cdot 10^{-2}}$ & $\mathrm{3.24 \cdot 10^{-2}}$ 
\\
$\mathrm{183 \pm6}$  & $ \mathrm{~(3.0 \pm 1.1) \cdot 10^{-2}}$ & $\mathrm{2.50 \cdot 10^{-2}}$ 
\\
$\mathrm{206 \pm15}$  & $ \mathrm{ ~(8.0 \pm 5.0) \cdot 10^{-3}}$ & $\mathrm{1.90 \cdot 10^{-2}}$ 
\\
$\mathrm{219 \pm7}$  & $ \mathrm{ ~(4.0 \pm 4.0) \cdot 10^{-3}}$ & $\mathrm{1.58 \cdot 10^{-2}}$ 
\\
$\mathrm{286 \pm11}$  & $ \mathrm{ ~(4.0 \pm 4.0) \cdot 10^{-3}}$ & $\mathrm{6.11 \cdot 10^{-3}}$ 
\\
$\mathrm{398 \pm 21}$ &$< \mathrm{4.2\cdot 10^{-3}}$ & $\mathrm{1.20 \cdot 10^{-3}}$
\\
$\mathrm{514 \pm23}$  &$< \mathrm{4.2\cdot 10^{-3}}$ & $\mathrm{2.39 \cdot 10^{-4}}$\\
\hline
\end{tabular*}
\end{table}

Surely, both the size and design main features of the MOSCAB detector do not make this detector suitable for monitoring the varying flux of background neutrons in the very extreme conditions existing inside LNGS underground laboratories. 
Nevertheless the general agreement between the detector counting rates and the expectations due to the environmental background
%, in the range 150 keV $\leq \xEc \leq $ 300 keV, 
corroborates the use of this detector to measure very low fluxes of fast neutrons in what we could define intrinsic background-free conditions.

\section{Conclusions}

MOSCAB bubble-chamber, originally designed for the direct observation of WIMPs in the spin-dependent channel, has been employed for the detection of fast neutrons using different configurations of the equipment and different amounts of sensitive liquid. 

First of all, it has been shown that the detector can be operated at metastability degrees such that the target liquid is sensitive only to neutrons. 
Actually, this occurs when the reduced superheat parameter SH is kept lower than 0.07, or the critical energy $\xEc$ is kept higher than nearly 200 keV, which approximately corresponds to a 2.5 MeV neutron energy threshold required for nucleation. 
In these conditions the residual internal background rate of the detector is lower than one event occurring every ten days and the steadiness of operation of the detector for long times of observation are extremely enhanced, as we demonstrated by running MOSCAB bubble-chamber during more than 4000 hours in the very clean environment of LNGS underground laboratories.

Subsequently, the detector response to fast neutrons, described through a bubble nucleation model assuming the existence of a critical deposition length of the energy released by the recoiled ions, has been investigated using a weak $^{241}$AmBe neutron source, which has allowed us to generate detection efficiency functions via Monte Carlo simulation and post-processing, whose validation has been performed experimentally.

Finally, thanks to the demonstrated reliability of the simulation procedure of the apparatus, we have displayed that MOSCAB bubble-chamber can be effectively used to measure the fast neutron activity of very weak n-sources, provided that the external background is almost negligible, as it happens inside the LNGS underground laboratories. 
On the other hand,  placed above ground without any additional shield, the same equipment could be exploited to monitor the cosmic ray variations through the neutron component of the Extensive Air Showers, as well as to detect the presence of any signal on top of the cosmic ray background. In particular, to study variations of the cosmic ray neutron flux whose energy spectrum extends to high energies, high neutron energy thresholds have to be set, which means that any other neutron source but cosmic rays will be practically absent, the information brought by the cosmic radiation being uncontaminated. Further investigations on this topic are scheduled to be conducted in the very next future.

\section{acknowledgements}
The authors wish to thank all the staff of the National Gran Sasso Laboratory for 
their constant support and cooperation during these years. 
Some of the scientists who imagined and realized the MOSCAB detector, Pietro Negri and Antonino Pullia, are not with us anymore. We are left with their memory and their teachings.
%\end{acknowledgements}

\end{document}